\documentclass[aps,prc,twocolumn,superscriptaddress]{revtex4-2}
\usepackage{lineno,isotope,mathrsfs}
\modulolinenumbers[5]
\usepackage{amsmath,bbold,amssymb,epsfig,bm,mathrsfs,feynmp}
\usepackage{color,slashed,exscale,multirow,soul,nicefrac,gensymb}
\usepackage{longtable,times,txfonts,xcolor,graphicx,tikz}
\usepackage[normalem]{ulem}
\usepackage[breaklinks,colorlinks,citecolor=blue]{hyperref}
\definecolor{red}{rgb}{0.8,0,0}
\definecolor{violet}{rgb}{0.4,0,0.4}
\definecolor{green}{rgb}{0,0.5,0.0}
\definecolor{navy}{rgb}{0.0,0.0,0.6}
\definecolor{orange}{rgb}{0.8,0.2,0.0}

\usepackage[normalem]{ulem}  

\newcommand{\bea}{\begin{eqnarray}}
\newcommand{\eea}{\end{eqnarray}}

\begin{document}
\title{Bayesian inferences on covariant density functionals 
from multimessenger astrophysical data: \\
Influences of parametrizations of density dependent couplings}
%
\author{Guo-Jun Wei}
\affiliation{Institute of Theoretical Physics, 
            Shanxi University, Taiyuan 030006, China}
\affiliation{School of Science, 
            Huzhou University, Huzhou 313000, China} 
\affiliation{School of Physical Science and Technology, 
             Southwest University, Chongqing 400715, China}       
\author{Jia-Jie Li}
\email{jiajieli@swu.edu.cn}           
\affiliation{School of Physical Science and Technology, 
             Southwest University, Chongqing 400715, China}         
\author{Armen Sedrakian}   
\email{sedrakian@fias.uni-frankfurt.de}         
\affiliation{Frankfurt Institute for Advanced Studies,
             D-60438 Frankfurt am Main, Germany}         
\affiliation{Institute of Theoretical Physics,
             University of Wroc\l{}aw, 50-204 Wroc\l{}aw, Poland} 
\author{Yong-Jia Wang}
\email{wangyongjia@zjhu.edu.cn}
\affiliation{School of Science, 
            Huzhou University, Huzhou 313000, China} 
\author{Qing-Feng Li}
\email{liqf@zjhu.edu.cn}
\affiliation{School of Science, 
             Huzhou University, Huzhou 313000, China} 
\affiliation{China Institute of Atomic Energy, 
             Beijing 102413, China}             
\author{Fu-Hu Liu} 
\affiliation{Institute of Theoretical Physics, 
            Shanxi University, Taiyuan 030006, China}
\begin{abstract}
Covariant density functionals have been successfully applied to 
the description of finite nuclei and dense nuclear matter. These
functionals are often constructed by introducing density dependence
into the nucleon–meson couplings, typically through functions that
depend only on the vector, i.e., proper baryon density. In this work, 
we employ a Bayesian framework to investigate how different 
parametrizations, characterized by distinct functional forms and by 
their dependencies on vector and scalar densities, affect the properties 
of dense matter and compact stars. Our analysis demonstrates that 
although all considered parametrizations yield broadly comparable 
inferences, the differences in the equation of state and the symmetry 
energy remain significant at suprasaturation densities, reflecting the
sensitivity to the chosen functional form of the density dependence.
We find that allowing the nuclear saturation properties in the
isoscalar channel, including the skewness coefficient $Q_{\rm sat}$, 
to be freely adjusted provides adequate flexibility for the current 
modeling of nuclear and neutron-star matter. In contrast, the isovector 
channel requires further refinement, with freedom extended at least up 
to the curvature coefficient $K_{\rm sym}$ to capture variations in 
the symmetry energy and particle composition at high densities. This work 
advances prior studies by implementing a rational-function parametrization 
of the density dependence, informed and constrained by multimessenger
astrophysical observations.
\end{abstract}
%
%
\date{\today}
\maketitle
%
\section{Introduction}
\label{sec:Intro}
Understanding the behavior of dense hadronic matter is one of the
central challenges in nuclear physics and astrophysics. The equation
of state (EOS) of nuclear matter serves as a crucial bridge between
the microscopic nucleon dynamics governed by nucleon–nucleon
interactions and the macroscopic properties of compact stars (CSs),
which are among the most extreme objects in the universe. With
densities up to $5$-$10$ times higher than those found in ordinary 
nuclei, CSs offer a unique setting for studying the EOS of dense 
matter under conditions that are impossible to achieve in terrestrial
laboratories; for reviews of the EOS of CSs, see
Refs.~\cite{Oertel:2017,Sedrakian:2023}.

Recent advances in astrophysical observations and terrestrial experiments 
have greatly enhanced our capacity to probe the nuclear EOS. These 
developments include: (a) constraints on CS deformability derived by 
LIGO–Virgo collaboration from gravitational waves emitted during binary 
neutron star mergers~\cite{LVScientific:2017,LVScientific:2018,LVScientific:2019,
LVScientific:2020a,LVScientific:2020b};
(b) radius measurements of several CSs obtained by the NICER x-ray
observatory~\cite{Riley:2019,Riley:2021,Miller:2019,Miller:2021,
Vinciguerra:2024,Choudhury:2024,Salmi:2024a,Salmi:2024b,Mauviard:2025}; 
and (c) measurements by the PREX and CREX
experiments~\cite{PREX:2021,CREX:2022,Reed:2021,Reinhard:2021} of the
neutron skin of nuclei in parity-violating electron scattering experiments 
that allowed the extraction of the slope of the symmetry energy. 
Theoretical modeling of PREX-2 experiment which includes also electric 
dipole polarizability measurements, suggests intermediate values of 
the slope of symmetry energy $54\pm 8$~MeV~\cite{Reinhard:2021}.
Additional valuable constraints are provided by ab initio studies
based on nucleon-nucleon potentials derived from the chiral effective
field theory ($\chi$EFT) framework. These include predictions for
neutron matter using realistic two- and three-nucleon interactions,
which remain valid within the $\chi$EFT setup up to approximately
$1.5$-$2$ times the saturation
density~\cite{Hebeler:2013,Lynn:2016,Drischler:2019,Keller:2023}. 
Similarly, ab initio calculations can be performed in the perturbative 
quantum chromodynamics (pQCD) regime at extremely high densities, around 
$40$ times the saturation density. Such calculations provide an effective
boundary condition for models that are extrapolated to the high-density
regime~\cite{Komoltsev:2022,Brandes:2023,Gorda:2023}.

The EOS of the dense hadronic phase in CSs can be constructed using a
variety of approaches, including density functional
methods~\cite{Traversi:2020,Malik:2022a,Malik:2022b,Beznogov:2024a,
Beznogov:2024b,Huang:2024,Lijj:2024c,Xiacj:2024},
meta-models~\cite{Margueron:2018,Zhang:2020,Tsang:2023,Margueron:2025},
and model-ag\-nos\-tic
descriptions~\cite{Raaijmakers:2019,Landry:2020,Legred:2021,Pang:2021,
Altiparmak:2022,Annala:2022,Chimanski:2022,Rutherford:2024,Fan:2024}. 
Among these, the covariant density functional (CDF)
theory~\cite{Nikolaus:1992,Vretenar:2005,Niksic:2011} — rooted in the quantum 
field theory of hadronic matter — provides a unified framework capable of
addressing the full range of nuclear data, from the nuclide chart to 
the physics of CSs~\cite{Oertel:2017,Piekarewicz:2020,Sedrakian:2023}. 
The class of CDF models is based on relativistic Lagrangians describing 
nucleons — constituents of dense hadronic matter — as interacting via 
meson exchange. They maintain key relativistic features, such as Lorentz
covariance, the Dirac structure of the self-energies, etc. Among the 
alternatives, we mention the point-coupling CDF models, which replace 
meson exchange interactions with four-point (contact) interactions between 
nucleons in various scalar and vector interaction 
channels~\cite{Nikolaus:1992,Burvenich:2002,Zhaopw:2010}. 
In recent years, the increasing availability of theoretical, observational, 
and experimental constraints has enabled the development of CDF-based
multiphysics Bayesian inference frameworks, which have emerged as a 
powerful tool for deriving and systematically refining constraints on
the CS and nuclear matter
EOS~\cite{Zhu:2023,Providencia:2023,Char:2023,Beznogov:2023,Malik:2023,
Salinas:2023,Mondal:2023,Zhou:2023,Parmar:2024,Scurto:2024,Xiacj:2024,
Lijj:2025a,Lijj:2025b,Char:2025,Cartaxo:2025}.

The most widely used CDF models, which rely on meson-exchange interaction 
between the nucleons, are generally categorized into two classes: (a) those
with constant meson–nucleon couplings that incorporate nonlinear meson
self-interactions and/or cross-interaction terms into the effective
Lagrangian, and (b) those that retain only linear couplings but introduce 
explicit density dependences in the couplings to account for medium 
modifications of the meson–nucleon vertices; for review see
Refs.~\cite{Oertel:2017,Piekarewicz:2020,Sedrakian:2023}.
While both types of CDF models show convergent behavior at low densities, 
where physics is constrained by well-understood properties of nuclear 
matter under laboratory conditions, their predictions regarding the behavior 
of matter under extreme conditions — such as at densities far beyond 
saturation — rely heavily on the underlying assumptions for the nonlinear 
terms in the Lagrangian or the functional dependence of the meson–nucleon
couplings.These models can be contrasted with the point-coupling models 
mentioned above~\cite{Nikolaus:1992,Burvenich:2002,Zhaopw:2010}, for which 
Bayesian analyses have likewise been carried out~\cite{Xiacj:2024}, adopting 
a flexible parametrization of density-dependent couplings, which led to 
quantitative constraints on the isoscalar and isovector coupling constants 
in such models. Our Bayesian approach allows us an alternative view and 
complementary check of high-density matter properties.

In the present work, we perform a comprehensive Bayesian analysis that
systematically compares meson-exchange based CDF models of the second type
(density-dependent CDFs) featuring different choices of the density —
either scalar or vector — entering the prescribed functional
dependence, while employing identical multiphysics constraints and a
consistent computational framework. It has long been recognized that
CDF models operate with two types of densities: the vector density 
and the scalar density~\cite{Fuchs:1995}. In most models, the
meson–nucleon couplings depend on the vector density with specific
functional forms characterizing this
dependence~\cite{Typel:1999,Hofmann:2001,Liqf:2004,Lalazissis:2005,
Gogelein:2008,Typel:2018,Malik:2022a,Malik:2022b,Lijj:2023b,Nanmz:2025}. 
Only recently has a dependence on the scalar density, or a mixture of 
the two densities, been explored in the description of finite
nuclei~\cite{Typel:2018} and CS matter~\cite{Lijj:2023b,Shrivastava:2025} 
using specific sets of parametrizations.  

Regarding the functional form of the density dependence, in the widely
adopted density-dependent (DD) CDF models, particularly the DD-ME
version~\cite{Typel:1999,Lalazissis:2005}, the isoscalar
meson–nuc\-le\-on couplings are described by two rational functions with
constraints imposed among their parameters, while an exponential function 
is used for the isovector meson. This formulation results in a model with 
seven free parameters, enabling adjustment of nuclear matter saturation 
properties up to the skewness coefficient $Q_{\rm sat}$ for symmetric 
nuclear matter and up to the slope coefficient $L_{\rm sym}$ for the symmetry
energy of isospin asymmetrical nuclear matter~\cite{Lijj:2019a,Lijj:2019b,Lijj:2024c}.
In the present work, we consider extensions and generalizations of the 
density-dependent functional forms of the meson–nucleon couplings to 
allow greater flexibility, with the goal of probing the influence of 
higher-order nuclear characteristic coefficients at saturation, defined 
through the familiar Taylor expansion of the energy density of nuclear 
matter around the saturation density and isospin symmetrical limits. 

The flexibility of a CDF can be increased by promoting higher-order 
(beyond $Q_{\rm sat}$ and $L_{\rm sym}$) coefficients to independent 
degrees of freedom to be determined through fits to data. While this 
added freedom allows the functional to explore a wider range of behaviors, 
it also introduces the risk of overfitting, especially when the available 
data are insufficient to constrain all parameters simultaneously. 
Our analysis shows that, given the current empirical and astrophysical 
constraints, allowing variations only in the low-order parameters 
already provides sufficient flexibility to reproduce the data at hand.

This naturally raises the question of how sensitive our results are to 
the specific choice of the CDF. Functionals with the same number of 
effective degrees of freedom as those considered here — such as the 
FSU model~\cite{Piekarewicz:2020,Salinas:2023} — are therefore expected 
to lead to qualitatively similar conclusions. In contrast, for models with 
fewer degrees of freedom, such as that employed in Ref.~\cite{Malik:2022a}, 
where the coefficients $K_{\rm sat}$ (incompressibility) and $Q_{\rm sat}$ 
(skewness) are strongly correlated by construction, the resulting distribution 
of $K_{\rm sat}$ is largely driven by that of $Q_{\rm sat}$. In such cases, 
the latter is primarily constrained by the two-solar-mass CS requirement, 
rather than by independent information on nuclear saturation properties.

For models with more degrees of freedom than those explored here, additional 
counterbalancing effects among parameters of different orders and in different 
isospin channels become possible. This may lead to noticeable changes in the 
inferred distributions of individual parameters. Nevertheless, this increased 
model flexibility does not alter the central conclusion of our study: with the 
presently available astrophysical data, low-order saturation parameters remain 
only weakly constrained. Improving these constraints will require either significantly 
more precise observational input or complementary information from laboratory 
experiments that directly probe nuclear matter near saturation density.

Thus, allowing a large number of free parameters would maximize flexibility but 
leave the problem underconstrained, whereas restricting the parameter set to a 
few well-determined quantities improves inferential power at the cost of increased 
model dependence through the assumed functional form of the density-dependent couplings.

This work is structured as follows: In Sec.~\ref{sec:Model} we discuss the model 
of CDF used in this work. In Sec.~\ref{sec:Bayesian} the Bayesian framework employed 
is outlined. Our results and their implications are presented in Sec.~\ref{sec:Results}. 
Section~\ref{sec:Conclusions} collects our conclusions and offers perspectives for 
the future work.

\section{Density-dependent CDF theory}
\label{sec:Model}
We now briefly review several key aspects of the CDF approach with
density-dependent couplings, putting the emphasis on the modifications
implemented in the functional forms of the density dependence.

\subsection{Effective Lagrangian and nucleon self-energy}
The Lagrangian of stellar matter with nucleonic degrees of freedom is 
given by the sum of the nucleonic, mesonic, and leptonic free 
Lagrangians, which can be found in 
Refs.~\cite{Oertel:2017,Piekarewicz:2020,Sedrakian:2023}, 
and the interaction Lagrangian, which reads
\begin{eqnarray}
\label{eq:Lagrangian}
\mathscr{L}_{\rm int} = \sum_{{\rm N} = n, p}
\bar{\psi}_{\rm N}\Big(
  g_{\sigma}\, \sigma
- g_{\omega}\,\gamma_\mu \omega^\mu 
- g_{\rho}\,\gamma_\mu \bm{\tau}_{\rm N} \cdot \bm{\rho}^\mu 
\Big)\,\psi_{\rm N}.
\end{eqnarray}
where $\psi_{\rm N}$ stands for the nucleonic [neutron $(n)$ and
proton $(p)$] fields, $\bm{\tau}_{\rm N}$ is the isospin vector, 
$g_{\rm m}$ ($\rm{m} = \sigma$, $\omega$ and $\rho$) are the 
meson-nucleon coupling strengths. The Lagrangian~\eqref{eq:Lagrangian} 
comprises the minimal set of interaction vertices necessary for 
a quantitative description of nuclear phenomena. The Dirac Hamiltonian 
for nucleons is given by
\begin{align}
\label{eq:Hamiltonian}
\mathscr{H} = \sum_{{\rm N} = n, p}\left[\bm{\alpha}\cdot\bm{p} + 
\beta\,\big(m_{\rm N} - 
\Sigma_{\rm S}\big) + \Sigma^{\rm N}_{\rm V}\right],
\end{align}
where  $\bm{\alpha}$ and $\beta$ are the well-known Dirac matrices, 
the first coupling the upper and lower components of the Dirac spinor, 
and the second, distinguishing between positive- and negative-energy 
states, $m_{\rm N}$ is the bare nucleon mass. 
The $\Sigma_{\rm S}$ and $\Sigma_{\rm V}^{\rm N}$ are the 
scalar and vector self-energies of an in-medium nucleon, and can 
be expressed as 
\begin{align}
\label{eq:Self_energy}
\Sigma_{\rm S} = g_\sigma\,\sigma + \Sigma_{\rm R}^{\rm s}, \qquad
\Sigma_{\rm V}^{\rm N} = g_\omega\,\omega + g_\rho\,\rho\,\tau_{\rm 3N} + 
\Sigma_{\rm R}^{\rm v}.
\end{align}
The rearrangement terms $\Sigma_{\rm R}^{\rm{s\,(v)}}$ arise from 
the dependence of the couplings on the scalar or vector densities, 
and are mandatory for maintaining the thermodynamic consistency of 
the model~\cite{Fuchs:1995,Typel:1999}. Note that in
Eq.~\eqref{eq:Self_energy} only the time-like components of the 
vector meson fields contribute, as required by rotational invariance 
and charge conservation.

In the mean-field approximation, the meson fields are replaced 
by their respective expectation values,
\begin{align}
\bar{\sigma} = \frac{1}{m_\sigma^2} g_\sigma \,n_{\rm s}, \qquad 
\bar{\omega} = \frac{1}{m_\omega^2} g_\omega \,n_{\rm v}, \qquad
\bar{\rho}   = \frac{1}{m_\rho^2} g_\rho\,n_{\rm v3},
\end{align}
where the scalar and vector densities at zero temperature are 
given by
\begin{subequations}\label{eq:Source_density}
\begin{align}
n_{\rm s} &\equiv \langle\bar{\psi}\psi\rangle = 
\sum_{{\rm N} = n,p} \frac{1}{\pi^2} \int^{k_{\rm F_N}}_0 
\frac{m_{\rm D,N}^\ast}{\sqrt{ 
k^2+m_{\rm D,N}^{\ast 2}} }k^2dk, \\
n_{\rm v} &\equiv \langle \psi^\dag\psi\rangle = 
\sum_{{\rm N} = n,p} \frac{1}{\pi^2} \int^{k_{\rm F_N}}_0 k^2dk = 
\sum_{{\rm N} = n,p} \frac{1}{3\pi^2}k^3_{\rm F_N},
\end{align}
\end{subequations}
with $k_{\rm F_N}$ being the neutron ($n$) or proton ($p$) Fermi
momentum, and $m_{\rm D, N}^\ast = m_{\rm N} - \Sigma_{\rm S}$ the 
Dirac effective mass. The difference between the scalar and vector 
(baryon) densities arises from the additional factor
$m_{\rm D,N}^\ast/(k^2 + m_{\rm D,N}^{\ast 2})^{1/2}$ in the
scalar-density integrand. This term reflects Lorentz contraction,
which suppresses the scalar contribution of highly relativistic
nucleons in the limit $k \gg m_{\rm D,N}^\ast$. Clearly, the scalar
density is always smaller than the vector density.

In numerical calculations, we assume $m_n = m_p\equiv m_{\rm N}$ fixed 
at the average bare mass of a nucleon; then, the nucleon Dirac mass is 
independent of the isospin. For lepton masses we take their vacuum values, 
whereas for the mesons we adopt masses close to their vacuum values, 
as given explicitly in Refs.~\cite{Lijj:2025a,Lijj:2025b}. The EOS and the 
composition of stellar matter are then obtained by combining the 
above relations with global charge neutrality and $\beta$-equilibrium 
conditions, which determine the relations among the chemical potentials 
of the various particle species~\cite{Lijj:2018a}.
We further match smoothly our core EOS to that of the crust EOS given in 
Refs.~\cite{Baym:1971a,Baym:1971b} at the crust-core transition density 
$n_{\rm sat}/2$. The integral parameters of a CS, specifically its mass, 
radius, and tidal deformability, are obtained by solving the 
Tolman-Oppenheimer-Volkoff equations~\cite{Tolman:1939,Oppenheimer:1939} 
along with the associated equations for tidal 
deformability~\cite{Flanagan:2008,Hinderer:2008}.

\subsection{Parametrizations of density-dependent couplings}
The density-dependent couplings $g_{\rm m}$ are the central 
quantities that determine the quality of a CDF model. 
They are usually written in the form:
\begin{align}
g_{\rm m}(n)=g_{\rm m}(n_{\rm ref})\,f_{\rm m}(r),
\end{align}
with constant values $g_{\rm m}(n_{\rm ref})$ at a reference density 
$n_{\rm{ref}}$ and in principle arbitrary functions $f_{\rm m}(r)$ 
that depend on the ratio $r=n/n_{\rm ref}$. Within the coupling vertices 
adopted in Lagrangian~\eqref{eq:Lagrangian} the (total) number density 
$n$ can be the vector or scalar density, and $n_{\rm ref}$, frequently, 
is the corresponding value at saturation density $n_{\rm sat}$. 

It is common to adopt for $f(r)$ either a rational (R)
function~\cite{Typel:1999,Lalazissis:2005}
\begin{equation}
\label{eq:rational_function}
f(r)=a\,\frac{1+b\,(r+d)^2}{1+c\,(r+d)^2},
\end{equation}
or an exponential (E) form,
\begin{equation}
\label{eq:exponential_function}
f(r)=\exp[-a\,(r-1)].
\end{equation}
These functions were originally introduced to reproduce results from
Dirac–Brueckner–Hartree–Fock calculations~\cite{vanDalen:2007}, with
all couplings required to decrease with increasing density. A notable
feature of the functional forms in Eqs.~\eqref{eq:rational_function}
and~\eqref{eq:exponential_function} is their well-behaved asymptotic
behavior as $r \rightarrow \infty$, where they approach either a
constant value (specifically $ab/c$) in the rational case, or vanish
in the exponential case.

In the widely used DD-ME
version~\cite{Typel:1999,Lalazissis:2005,Lijj:2023b}, the couplings
$g_{\rm m}$ depend on the vector (V) density $n_{\rm v}$. For the
isoscalar mesons $\sigma$ and $\omega$, the rational
function~\eqref{eq:rational_function} is employed, but with additional
constraints. Specifically, for each meson, the function is normalized
at saturation $r = 1$ and its curvature vanishes at 
$r = 0$~\cite{Typel:1999,Lalazissis:2005},
\begin{align}
\label{eq:coupling_constraints_1}
f(1)=1, \quad f^{\prime\prime}(0)=0,
\end{align}
and  the two mesons share the constraint
\begin{align}
\label{eq:coupling_constraints_2}
f^{\prime\prime}_{\sigma}(1)=f^{\prime\prime}_{\omega}(1).
\end{align}
These restrictions reduce the number of independent parameters 
to three, i.e., $a_\sigma$, $d_\sigma$, and
$d_\omega$~\cite{Lijj:2024c,Lijj:2025a}.

To extend this modeling framework, we remove the
condition~\eqref{eq:coupling_constraints_2}, thereby increasing the
number of independent parameters to four, specifically, $b_\sigma$,
$d_\sigma$, $b_\omega$, and $d_\omega$. This modification provides
additional flexibility in describing the high-density behavior of
symmetric nuclear matter.

For the isovector $\rho$-meson, the exponential
form~\eqref{eq:exponential_function} with a single parameter
$a_{\rho}$ has been commonly
used~\cite{Typel:1999,Lalazissis:2005,Lijj:2023b}. In this case, 
the coupling decreases rapidly with density, approaching zero at 
high densities.

In the present work, we explore an alternative parametrization of
the $\rho$-meson coupling by using  the rational
function~\eqref{eq:rational_function} with the following three 
conditions imposed: 
\begin{align}
\label{eq:coupling_constraints_3}
f(1)=1, \quad f^{\prime}(0) = 0, \quad
f^{\prime}(1)/f(1) = f^{\prime\prime}(1)/f^{\prime}(1).
\end{align}
This leads to a function with one independent parameter (i.e., $c_{\rho}$), 
constructed to mimic the low-density behavior of the exponential form. 
Furthermore, if we also apply the conditions given
in~\eqref{eq:coupling_constraints_1}, the resulting parametrization
contains two independent parameters, $b_\rho$ and $d_\rho$,
providing additional flexibility for describing the symmetry energy 
at high densities.
\begin{table}[tb]
\centering
\caption{
The CDF models that are explored in this work. For each model, 
we specify the type of density dependence, the functional form, 
condition applied to the isoscalar $\sigma$ and $\omega$ mesons 
and the isovector $\rho$ meson, and the total number of independent 
parameters of the model in the last column.
}
\setlength{\tabcolsep}{7.8pt}
\label{tab:CDF_models}
\begin{tabular}{ccccc}
\hline\hline
Model & Type & $f_\sigma$, $f_\omega$   & $f_\rho$ & No. \\
\hline
VRE   & V  & R, Eqs.~\eqref{eq:coupling_constraints_1} and~\eqref{eq:coupling_constraints_2} & E & 7 \\
[0.3em]
MRE   & M  & R, Eqs.~\eqref{eq:coupling_constraints_1} and~\eqref{eq:coupling_constraints_2} & E & 7 \\
MRE2  & M  & R, Eq.~\eqref{eq:coupling_constraints_1}  & E & 8 \\
[0.3em]
VRR   & V  & R, Eqs.~\eqref{eq:coupling_constraints_1} and~\eqref{eq:coupling_constraints_2} & R, Eq.~\eqref{eq:coupling_constraints_3} & 7 \\
VRR2  & V  & R, Eqs.~\eqref{eq:coupling_constraints_1} and~\eqref{eq:coupling_constraints_2} & R, Eq.~\eqref{eq:coupling_constraints_1} & 8 \\
\hline\hline
\end{tabular}
\end{table}
%

As alternative parametrizations, one can also construct couplings 
that depend on the total scalar (S) density $n_{\rm s}$, or a mixed (M)
density dependence on $n_{\rm v}$ and $n_{\rm s}$ simultaneously,
where the vector meson couplings $g_\omega$ and $g_\rho$ depend on
$n_{\rm v}$, and the scalar meson coupling $g_\sigma$ depends on
$n_{\rm s}$, as implemented in Refs.~\cite{Typel:2018,Lijj:2023b}.

In the present work, we explore several combinations for the choice of
the density entering the argument $r$, the functional form of $f(r)$,
and the conditions imposed on it. Table~\ref{tab:CDF_models} summarizes 
these options, using a three-letter abbreviation for identification. 
The abbreviation indicates, respectively, the type of density used, 
the functional forms for the two isoscalar mesons ($\sigma$ and $\omega$), 
and the functional form for the isovector meson ($\rho$). For example, 
the DD-ME version recently explored in 
Refs.~\cite{Lijj:2024c,Lijj:2025a,Lijj:2025b,Char:2025,Cartaxo:2025}
is abbreviated as ``VRE'', indicating a vector density for normalization, 
a rational function in the isoscalar channel, and an exponential function 
in the isovector channel. The conditions imposed on the $f(r)$ 
functions are made explicit in Table~\ref{tab:CDF_models}.

%
\begin{figure*}[tb]
\centering
\includegraphics[width = 0.99\textwidth]{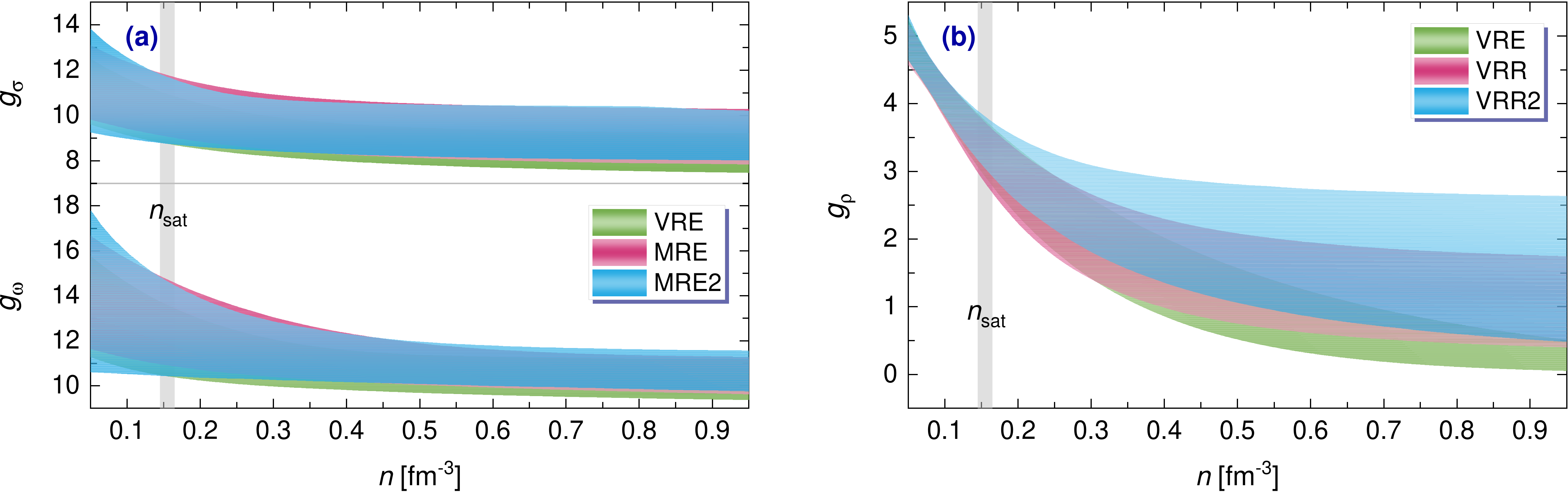}
\caption{
The posterior confidence regions (95.4\% CI) for meson-nucleon 
couplings $g_\sigma$, $g_\omega$ and $g_\rho$ obtained using CDFs 
with different parametrizations for the density dependence.
The vertical bands correspond to the saturation density. 
}
\label{fig:Couplings}
\end{figure*}
%

A further key step is to relate the CDF parameters introduced above 
to the nuclear characteristic quantities at saturation density.
For small isospin asymmetry $\delta = (n_n - n_p)/n$, the energy density 
is decomposed as
\begin{equation}
\label{eq:E_expansion}
\varepsilon\,(n,\delta) \simeq n\,E_{\rm SNM}(n) + n\,E_{\rm sym}(n)\,\delta^{2},
\end{equation}
where $E_{\rm SNM}$ and $E_{\rm sym}$ denote the symmetric-matter and 
symmetry-energy contributions, respectively. These functions are then 
expanded around saturation density using the dimensionless parameter
$
\chi \equiv (n-n_{\rm sat})/{3\,n_{\rm sat}},
$
in the generic form
\begin{equation}
\label{eq:Taylor_expansion}
E_{\rm SNM}(n)=\sum_{i=0}^{4} u_i\,\chi^{i}, \qquad
E_{\rm sym}(n)=\sum_{i=0}^{4} w_i\,\chi^{i}.
\end{equation}
The coefficients $u_i$ encode the standard empirical characteristics 
of symmetric nuclear matter:
\begin{equation}
E_{\rm sat} = u_0,\quad
K_{\rm sat} = 2\,u_2,\quad
Q_{\rm sat} = 6\,u_3,\quad
Z_{\rm sat} = 24\,u_4,\nonumber
\end{equation}
corresponding to the saturation energy, incompressibility, skewness, 
and kurtosis, respectively. Similarly, the coefficients $w_i$ determine 
the symmetry-energy characteristics:
\begin{gather}
J_{\rm sym} = w_0,\quad
L_{\rm sym} = w_1,\nonumber \\
K_{\rm sym} = 2\,w_2,\quad
Q_{\rm sym} = 6\,w_3,\quad
Z_{\rm sym} = 24\,w_4,\nonumber
\end{gather}
representing the symmetry energy at saturation, its slope, curvature, 
skewness, and kurtosis.

\section{Inference framework and constraints}
\label{sec:Bayesian}
To ensure this presentation is self-contained, we briefly review 
the Bayesian inference setup of Refs.~\cite{Lijj:2024c,Lijj:2025a}. 
The framework incorporates both multi-physics constraints from 
astrophysical observations and terrestrial experiments, as well as 
microscopic theory computations, to inform a CDF model of 
dense matter. 

Specifically, for the constraints sensitive to the low-density 
re\-gime of nucleonic EOS, we consider: (i) a collection of nuclear 
matter characteristics at saturation
$\big\{\,m^\ast_{\rm{D}},\,n_{\rm{sat}},\, E_{\rm{sat}},\,K_{\rm{sat}},\,J_{\rm{sym}}\,\big\}$, 
which are well constrained by the low-energy nuclear phenomena~\cite{Lijj:2025a};
(ii) the energy per particle and pressure for pure neutron matter
derived from $\chi$EFT interactions~\cite{Hebeler:2013} within their
validity range, i.e., densities below $1.5$-$2$ times the saturation density.
For the astrophysical constraints, we incorporate: (i) the mass
measurement of massive pulsars PSR J0348+0432~\cite{Antoniadis:2013};
(ii) the tidal deformabilities obtained for the binary
neutron star mergers GW170817 and GW190425, by the LIGO-Virgo
Collaboration~\cite{LVScientific:2017,LVScientific:2020a}; (iii) the
NICER's simultaneous mass and radius estimates for four millisecond
pulsars -- the $\sim 2.0\,M_\odot$ pulsar PSR
J0740+6620~\cite{Salmi:2024a}, the two canonical mass
$\sim 1.4\,M_\odot$ objects PSR J0030+0451~\cite{Vinciguerra:2024} and
J0437-4715~\cite{Choudhury:2024}, and the light $\sim 1.0\,M_\odot$ 
pulsar PSR J1231-1411~\cite{Salmi:2024b}. The constructions of likelihood 
functions for each constraint are detailed in Refs.~\cite{Lijj:2024c,Lijj:2025a}.

The Bayesian parameters of the CDF model in our analysis include the
three coupling strengths ($g_{\rm m}$, with
${\rm m} = \sigma,\,\omega,\,\rho$) at saturation density $n_{\rm sat}$,
along with four or five parameters in the functions $f_{\rm m}$ that
govern their density dependence. Uniform prior distributions are
assigned to these parameters over carefully chosen intervals: wide
enough to avoid biasing the posterior distributions, yet not so broad
as to impede the convergence of the Monte Carlo algorithm.

\section{Results and implications}
\label{sec:Results}
To assess the individual impact of each channel of the CDF model on
the Bayesian inference, we held the density-dependence of the function
$f(r)$ fixed in one channel (isoscalar or isovector) and varying it 
in the other, applying identical multi-messenger constraints
in all cases.

In Fig.~\ref{fig:Couplings}, we show the Bayesian inference of the
density dependence of the meson-nucleon couplings at 95.4\% confidence
interval (CI) for each of the CDF models listed in
Table~\ref{tab:CDF_models}. In panel (a) the models VRE, MRE, and MRE2
feature different rational functions for the $\sigma$- and
$\omega$-meson couplings, but share the same exponential function
$\rho$-meson coupling, as does also the VRE model shown in panel (b).
By contrast, in panel (b), the models VRE, VRR, and VRR2 feature either
rational or exponential functions for the $\rho$-meson coupling, but
share same rational function for the $\sigma$- and $\omega$-meson
couplings, as does also the VRE model shown in panel (a). 

From panel (a) of Fig.~\ref{fig:Couplings}, one sees that the couplings 
$g_\sigma$ and $g_\omega$ exhibit a self-similar density dependence.
This reflects their complementary roles in Eq.~\eqref{eq:Self_energy}: 
$g_\sigma$ controls the attractive scalar self-energy $\Sigma_{\rm S}$ 
through the $\sigma$-meson field, while $g_\omega$ governs the repulsive 
vector self-energy $\Sigma_{\rm V}^{\rm N}$ through the $\omega$-meson 
field. The delicate balance between $\Sigma_{\rm S}$ and 
$\Sigma_{\rm V}^{\rm N}$ determines the EOS of symmetric nuclear matter.

%
\begin{figure*}[tb]
\centering
\includegraphics[width = 0.99\textwidth]{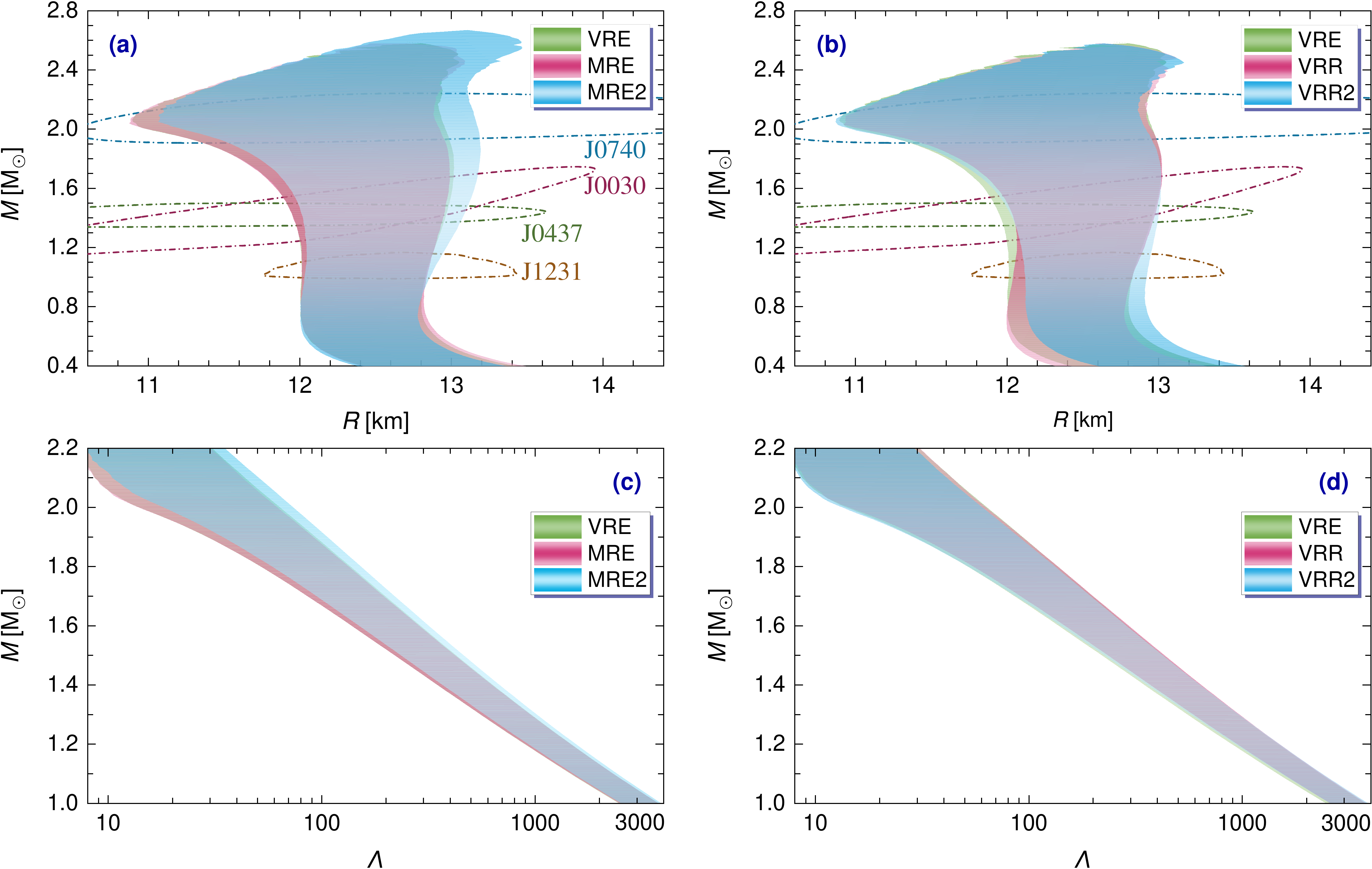}
\caption{ Mass–radius and mass-tidal deformability diagrams 
  for CSs, with elliptical contours indicating the 95.4\% CI 
  regions of mass–radius estimates for four pulsars from NICER 
  observations. Left panels (a and c) show the posterior 95.4\% CI 
  regions obtained using CDFs with different isoscalar coupling 
  parametrizations, while the right panels (b and d) presents the
  results for variations in the isovector coupling.}
\label{fig:MR_NS}
\end{figure*}
%

The range of coupling values spanned by each model determines its
flexibility in simultaneously describing different constraints. 
In panel (b) of Fig.~\ref{fig:Couplings}, the rational function
subject to conditions~\eqref{eq:coupling_constraints_3} reproduces 
the modeling of the $g_\rho$ coupling with an exponential function at
densities up to $2\,n_{\rm sat}$, but deviates significantly at higher
densities. Under conditions~\eqref{eq:coupling_constraints_2}, the 
rational function allows much larger $g_\rho$ values at high densities. 
Nonetheless, the narrow range and consistency of all three models at 
and below $n_{\rm sat}$ are a consequence of the $\chi$EFT constraint 
for pure neutron matter~\cite{Lijj:2024c,Lijj:2025a}.

%
\begin{figure*}[tb]
\centering
\includegraphics[width = 0.99\textwidth]{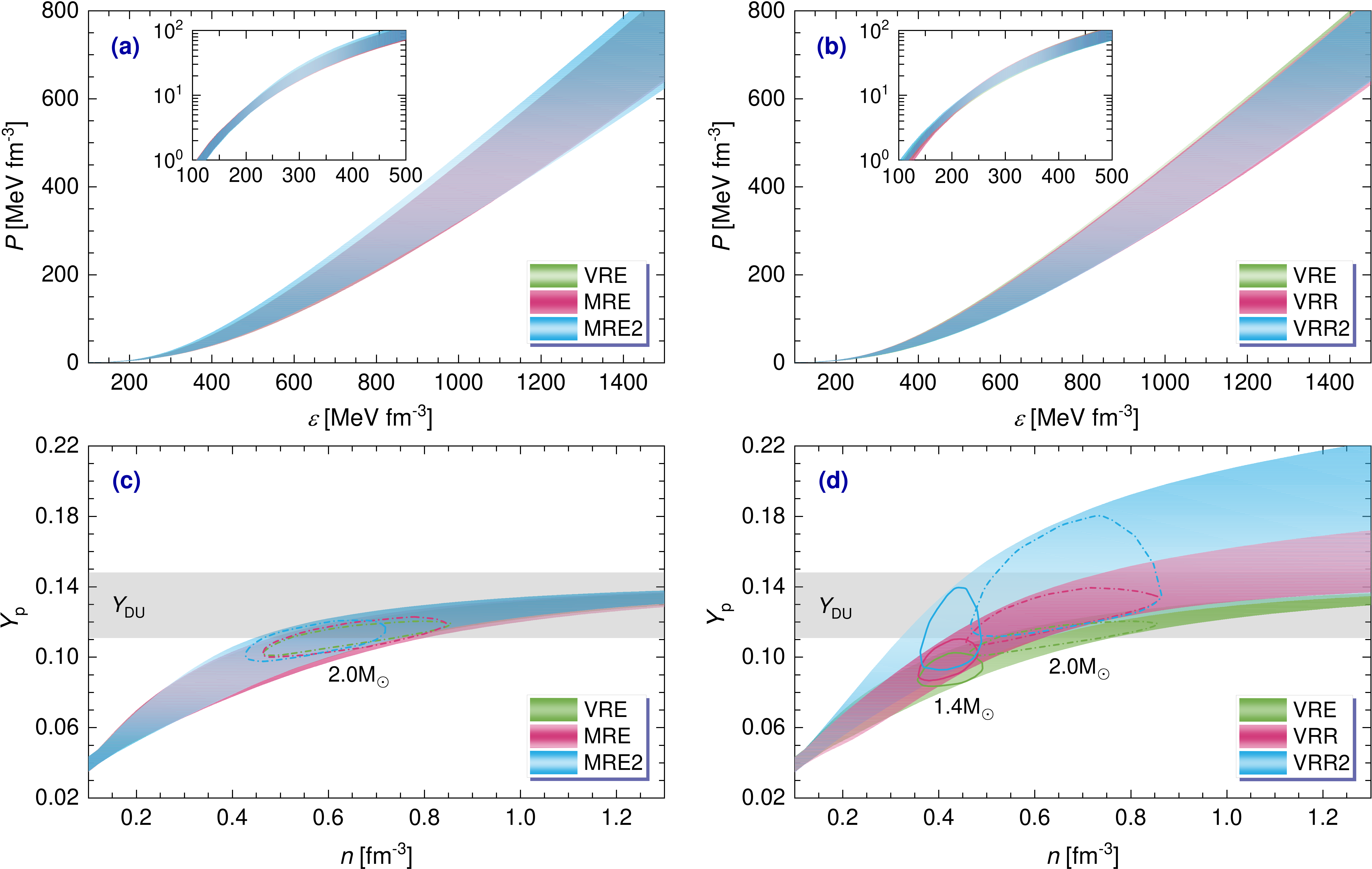}
\caption{The influences of parametrizations of density-dependent
  couplings on the $\beta$-equilibrium EOS and its particle
  composition characterized by proton fraction $Y_{\rm p}$. 
  Left panels (a and c) show the posterior confidence regions (95.4\% CI) 
  obtained using CDFs with different parametrizations of isoscalar couplings,
  while the right panels (b and d) show the corresponding results for
  variations in the isovector coupling. In the upper panels, the 
  insets magnify the low-density regime. In the lower panels, the
  contours show the corresponding distributions of the respective 
  1.4 and 2.0~$M_{\odot}$ stars, and the horizontal bands labeled
  $Y_{\rm DU}$ indicate the admissible proton fraction thresholds
  for the onset of nuclear direct Urca (DU) cooling process. }
\label{fig:EOS_NS}
\end{figure*}
%
\begin{figure*}[tb]
\centering
\includegraphics[width = 0.99\textwidth]{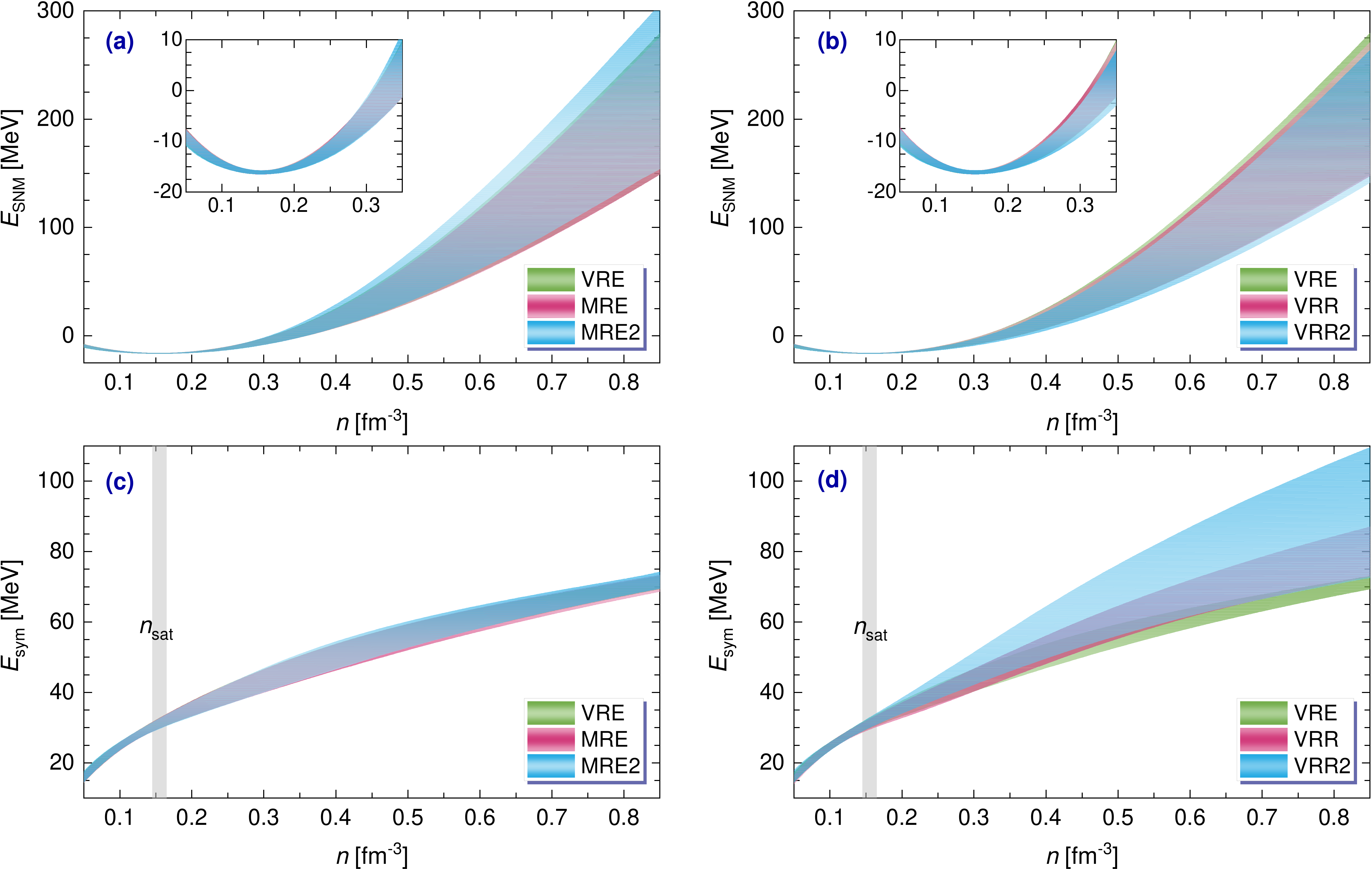}
\caption{
The influences of parametrizations of density-dependent couplings
on the energy per particle $E_{\rm SNM}$ of symmetric nuclear matter
and symmetry energy $E_{\rm sym}$. Left panels (a and c) show the 
posterior confidence regions (95.4\% CI) obtained using CDFs with 
different parametrizations of isoscalar couplings, while the right 
panels (b and d) show the results for variations in the isovector 
coupling. In the upper panels, the insets magnify the low-density regime.
}
\label{fig:EOS_NM}
\end{figure*}
%
\begin{figure*}[tb]
\centering
\includegraphics[width = 0.99\textwidth]{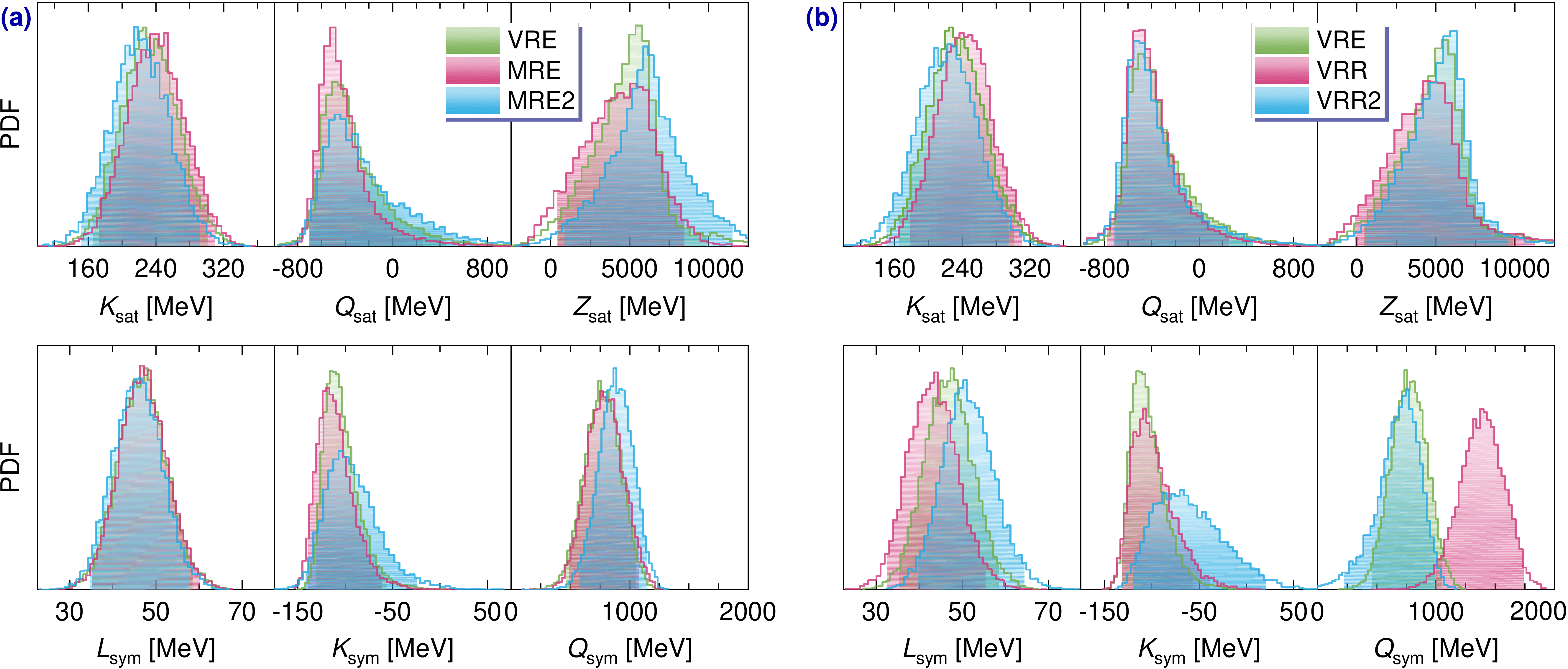}
\caption{
The posterior distributions for the nuclear matter characteristic 
coefficients at saturation density, where shaded regions correspond 
to the 95.4\% CI. Left panel (a) shows the posteriors obtained 
using CDFs with different parametrizations of isoscalar couplings; 
while the right panel (b) shows the results for variations in the 
isovector coupling.
}
\label{fig:NM_Chara}
\end{figure*}
%

The bulk properties of CSs -- such as maximum mass, radii, tidal 
deformabilities, and moments of inertia -- are fully determined
by the distributions of pressure and energy density throughout the
star, as prescribed by the dense matter EOS. Our inferences for the
mass-radius and mass-tidal deformability diagrams of CSs and various 
thermodynamic quantities of dense matter are shown in 
Fig.~\ref{fig:MR_NS}–\ref{fig:NM_Chara}.
Specifically, we show posterior 95.4\% confidence regions, whereby we
distinguish the results obtained using CDFs with different
parametrizations of the isoscalar couplings (left panels) and those
with variations in the isovector coupling (right panels). 
The new and updated mass-radius estimates for four pulsars obtained 
by the NICER 
collaboration~\cite{Salmi:2024a,Vinciguerra:2024,Choudhury:2024,Salmi:2024b} 
are also shown in Fig.~\ref{fig:MR_NS} for the sake of reference. 
The posterior distributions for $M$-$\Lambda$ exhibit similar trends 
to the $M$-$R$ ones due to the universal {\it $C$-Love} relation (see Refs.~\cite{Sedrakian:2023,Lijj:2023a} and references therein). 
A generic observation that follows from the analysis of results in 
Figs.~\ref{fig:EOS_NS}–\ref{fig:NM_Chara}, which will be given below, 
is that the interplay between isoscalar and isovector channels has a 
much weaker influence on the thermodynamics of dense matter than the 
effects of different parametrizations within each respective isospin
channel. Below, we discuss the impact of specific parametrizations 
in detail.

\subsection{Influences of parametrizations of isoscalar couplings}
We begin by examining the impact of different parametrizations of 
the isoscalar couplings ($g_\sigma$ and $g_\omega$) on CS and
dense-matter properties. To this end, we compare the posterior
distributions obtained with the VRE model against those from the newly
introduced MRE and VRE2 models. Figure~\ref{fig:MR_NS} displays the
posteriors for the mass–radius relation, while Fig.~\ref{fig:EOS_NS}
shows the corresponding distributions of the $\beta$-equilibrium EOS,
together with the particle composition characterized by the proton
fraction $Y_{\rm p}$, which is fully determined by the density dependence 
of the symmetry energy.

In both the VRE and MRE models, the number of parameters is sufficient
to independently adjust the nuclear characteristics at saturation
density up to and including the isoscalar skewness coefficient
$Q_{\rm sat}$, and up to the isovector symmetry-energy slope 
coefficient $L_{\rm sym}$~\cite{Lijj:2019b,Lijj:2023b}. The left 
panels of Figs.~\ref{fig:MR_NS} and~\ref{fig:EOS_NS} demonstrate that 
the 95.4\% CI posteriors for the VRE and MRE models are nearly identical. 
This consistency is further supported by the behavior of the symmetric
nuclear matter EOS and the symmetry energy shown in the left panels of
Fig.~\ref{fig:EOS_NM}.

The characteristic coefficients of nuclear matter, which quantify its
properties around saturation density, are summarized in
Fig.~\ref{fig:NM_Chara}. We omit the 0th-order coefficients, the
saturation energy $E_{\rm sat}$, and the symmetry energy at saturation
$J_{\rm sym}$, because they are well constrained and consistent across
all models considered in this analysis. For both the VRE and MRE
models, the posterior distributions of each displayed coefficient behave
similarly and exhibit comparable peak locations. It is important to
note, however, that the higher-order coefficients, specifically
$Z_{\rm sat}$, $K_{\rm sym}$, and $Q_{\rm sym}$, are not tunable
within the available parametrizations. Instead, their values emerge 
as predictions of the respective models~\citep{Lijj:2019b,Lijj:2023b}.

The MRE2 model, which includes one additional independent parameter 
compared with the MRE model, predicts larger radii by approximately
$0.1$-$0.3$~km for CSs with masses $M \gtrsim 1.4\,M_{\odot}$, along
with slightly higher maximum masses $M_{\rm max}$ by
$\sim 0.1\,M_{\odot}$, as shown in panel (a) of Fig.~\ref{fig:MR_NS}.
This is because the pulsar PSR J0740+6620, which favors a larger radius,
reforces the selection of EOS that are stiffer at high densities.
Correspondingly, the left panels of Fig.~\ref{fig:EOS_NS} indicate
that it yields a stiffer $\beta$-equilibrium EOS in the relevant
density regime, although the isovector channel, represented by the
proton fraction $Y_{\rm p}$, remains unaffected. This increased stiffness 
is also present in the EOS of symmetric nuclear matter at densities
$n \gtrsim 3\,n_{\rm sat}$, as seen in the left panels of
Fig.~\ref{fig:EOS_NM}.  In a recent study, Ref.~\cite{Cartaxo:2025}
compared the computations performed with the VRE model with
those that employed simplified functions given by
Eq.~\eqref{eq:exponential_function} for the isoscalar couplings, that
contain fewer adjustable parameters and therefore have reduced
flexibility. This comparison showed that the model with simple 
exponential dependence predicts larger radii for CSs than the VRE
model -- a trend that is opposite to our findings. This discrepancy
may be attributed to the fact that for the simple exponential
representation of the density dependence, the coefficient $Q_{\rm sat}$
is not independent, but is constrained by the fits of lower order
nuclear characteristics to a large range of values~\cite{Malik:2022a}.

Because small variations in the incompressibility $K_{\rm sat}$
(order-2) have a negligible impact on the high-density
EOS~\cite{Margueron:2018,Lijj:2025a}, the stiffness of the EOS in the 
MRE2 model arises primarily from the substantially broader distributions 
of the skewness coefficient $Q_{\rm sat}$ (order-3) and the kurtosis 
coefficient $Z_{\rm sat}$ (order-4). As shown in panel (a) of 
Fig.~\ref{fig:NM_Chara}, these distributions are shifted toward
larger values compared with the much narrower distributions obtained
with the MRE model. Interestingly, all three models predict
consistent values for the lower bound (95.4\% CI) of $Q_{\rm sat}$.

As is well known, the higher the order of a characteristic coefficient,
the farther from $n_{\rm sat}$ its influence becomes visible in the
behavior of the EOS and related thermodynamic quantities. At the same
time, the assessment of the impact of each specific coefficient is 
affected by the increasing uncertainty that accompanies higher-order 
terms. This behavior naturally follows from the Taylor 
expansion~\eqref{eq:Taylor_expansion}: contributions from higher-order 
coefficients carry progressively smaller weights and, therefore, 
large variations are needed to cover a prescribed range of priors.

For the MRE2 model, we further examined the correlations between
$Q_{\rm sat}$ and key CS observables, such as the maximum mass and the 
radius of a $2.0\,M_{\odot}$ star. These correlations are relatively 
strong, with Pearson coefficients of $r_{\rm p} \approx 0.7$-$0.8$. 
In contrast, the correlations involving $Z_{\rm sat}$ are weak, with 
$r_{\rm p} \approx 0.3$. Therefore, the skewness coefficient 
$Q_{\rm sat}$ emerges as the most important characteristics of the 
isoscalar part 
of the EOS above $\sim 3\,n_{\rm sat}$ within the density regime 
relevant for CSs.

\subsection{Influences of parametrizations of isovector coupling}
We now evaluate the influence of different parametrizations of the
isovector coupling ($g_\rho$) on the properties of CSs and dense
nuclear matter.  We compare the posteriors obtained from the VRE 
model with those from the newly introduced VRR and VRR2 models.

As shown in panel (b) of Fig.~\ref{fig:MR_NS}, the choice of $g_\rho$ 
parametrization has only a modest effect on the global properties of
CSs. The most noticeable impact occurs for low- to intermediate-mass 
stars, where the predicted radii vary by approximately 0.1~km in either 
direction, depending on the model. This behavior arises from modifications 
to the $\beta$-equilibrium EOS at low densities ($n \lesssim 2\,n_{\rm sat}$), 
as illustrated in panel (b) of Fig.~\ref{fig:EOS_NS}. These differences 
are closely tied to small variations in the symmetry-energy slope coefficient 
$L_{\rm sym}$ among the models. In the high-density regime, the posteriors 
from all three models overlap substantially, despite significant differences 
in the predicted proton fractions. This indicates that the rational-function 
para\-metrizations of $g_\rho$ introduced in the VRR and VRR2 models 
primarily modify the high-density behavior of the particle composition, 
permitting larger proton fractions in the stellar core—a notable departure 
from the original exponential form used in VRE. These features are 
qualitatively consistent with the findings of Ref.~\cite{Cartaxo:2025}, 
where an extended exponential function, 
$f_\rho\,(r) = y\,\exp[-a\,(r-1)] + (1-y)$ with $y$ as an additional 
parameter was employed.
 
The consistency of the results for the $\beta$-equilibrium matter EOS
across the three models in the high-density regime can be understood
through the decomposition of the energy density of the isospin-asymmetric 
matter Eq.~\eqref{eq:E_expansion}.
As shown in the right panels of Fig.~\ref{fig:EOS_NM}, a stiffer 
symmetric-matter EOS may be accompanied by a softer symmetry energy. 
The latter reduces the proton fraction in dense matter, as illustrated 
in panel (d) of Fig.~\ref{fig:EOS_NS}, leading to a larger isospin 
asymmetry $\delta$. As a result, the interplay between the isoscalar 
and isovector channels produces broadly comparable posterior distributions
for the pressure and energy density.

These findings indicate that the employed astrophysical constraints 
in the form of integral parameters of CS (mass, radius, etc.) are only 
weakly sensitive to the isovector channel of the EOS models. In addition,
extracting the symmetry energy and its density dependence from the
$\beta$-equilibrium EOS carries substantial uncertainties, which are
further amplified by model-specific treatments of the isovector channel. 
This is particularly true for the VRE and VRR models, both of which 
contain two independent isovector parameters. It should also be
emphasized that the dominant uncertainties arise in the high-density
regime, whereas the low-density behavior is well constrained by
$\chi$EFT calculations~\cite{Lijj:2024c}.

In the right panel of Fig.~\ref{fig:NM_Chara}, all three models
exhibit comparable ranges for the slope coefficient $L_{\rm sym}$
(order-1), with median values differing by no more than about 5
MeV. The higher-order coefficients -- the curvature $K_{\rm sym}$ (order-2)
and the skewness $Q_{\rm sym}$ (order-3) -- therefore provide the main
distinctions between the models. However, in the VRE and VRR models,
both $K_{\rm sym}$ and $Q_{\rm sym}$ arise as predictions, since the
two independent parameters in the $g_\rho$ coupling are already
tightly constrained by the values of $J_{\rm sym}$ and $L_{\rm
  sym}$. In contrast, the VRR2 model introduces a third independent
parameter in the $g_\rho$ coupling, allowing additional freedom to
vary $K_{\rm sym}$. Indeed, we find that the correlation between
$L_{\rm sym}$ and $K_{\rm sym}$ is negligible, with a Pearson
coefficient of $r_{\rm p} \approx 0.1$

The VRE and VRR models yield similar posterior distributions for the 
curvature coefficient $K_{\rm sym}$, but noticeably different distributions 
for the skewness coefficient $Q_{\rm sym}$. Conversely, the VRE and VRR2 
models produce comparable distributions for $Q_{\rm sym}$, while their 
predictions for $K_{\rm sym}$ differ substantially. Taken together with 
the symmetry energy shown in the lower panels of Fig.~\ref{fig:EOS_NM}, 
these results demonstrate that $K_{\rm sym}$ is the most influential 
nuclear coefficient governing the density dependence of the symmetry energy
beyond $2\,n_{\rm sat}$. The coefficient $Q_{\rm sym}$ also contributes
significantly, particularly at densities above $3\,n_{\rm sat}$.

Constraining the nuclear symmetry energy, particularly its high-density 
behavior, requires complementary strategies to those employed here. 
Observational data on the surface luminosities of cooling CSs offer 
valuable insight into the composition of dense matter and thus into the 
high-density symmetry energy~\cite{Blaschke:2004,Page:2006,Page:2011,Potekhin:2015}. 
It is therefore instructive to examine in detail the predicted proton 
fractions inside CSs across different models, especially regarding the 
possible onset of the nucleonic direct Urca (DU) process, which, if 
allowed, would lead to rapid stellar cooling~\cite{Lattimer:1991,Prakash:1992,Schaab:1996gd}.
Cooling simulations generally indicate that the DU process should not
operate in typical CSs with masses
$M < 1.5\,M_{\odot}$~\cite{Blaschke:2004,Popov:2006}, 
while various studies have proposed threshold masses in the range of
$1.6$-$1.8\,M_{\odot}$ for its
onset~\cite{Brown:2009,Beznogov:2014,Beznogov:2015,Brown:2018,Beloin:2019}. 
More recently, magneto-thermal simulations of young, cold CSs
suggest that viable dense-matter EOSs must permit fast cooling 
at least within certain mass intervals~\cite{Marino:2024}.

In the lower panels of Fig.~\ref{fig:EOS_NS}, we also show contours
corresponding to CS configurations with 1.4 and $2.0\,M_{\odot}$. 
The horizontal bands indicate the nucleonic DU 
threshold~\cite{Klahn:2006}: the lower limit of 11.1\% is obtained
from a $\mu$-free model, while the upper limit assumes massless
muons—though this applies primarily at high densities. It is evident
that the VRE model strongly disfavors the DU process in CSs with
$M \lesssim 2.0\,M_{\odot}$, consistent with
Refs.~\cite{Lijj:2024c,Cartaxo:2025}. In contrast, the VRR2 model
permits the DU process even in stars with masses as low as
$1.4\,M_{\odot}$, while the VRR model yields intermediate
predictions. These results suggest that the present rational-function
parametrizations of the $\rho$-meson coupling warrant further
applications in studies of cold CSs and, by extension, also supernova
and binary neutron star merger matter.

\section{Conclusions}
\label{sec:Conclusions}
In this work, we performed a systematic comparison of Ba\-yes\-ian 
inference outcomes for CDF-type EOS of dense matter, with a 
particular focus on the influences of different functional forms 
describing vector- or scalar-density dependence beyond the standard 
assumptions. All CDF models were constrained by the same set of 
nuclear and astrophysical observables, using identical methods to 
construct the likelihood functions. The prior intervals for the CDF 
parameters were chosen to be sufficiently wide to avoid  biasing 
the posterior distributions.

Our analysis evaluated the resulting differences in inferences by
examining the properties of CSs and dense matter, along with the 
particle composition. We found that CDF models featuring the same
functional form of density dependence -- but differing in the type of
density (vector or scalar) employed in the density-parametrization --
yield nearly identical posterior results for the global properties of
CSs (such as maximum mass, radii, and tidal deformabilities) and dense
matter (including the EOS, nuclear characteristic coefficients, and
particle composition) with the 95.4\% CI essentially overlapping.

Extending the functional form of the density dependence in the 
isoscalar channel to achieve greater flexibility primarily affects
dense matter at high densities: while its impact on global CS properties 
are modest, the proton fraction in massive stars can be significantly 
unaltered. Consequently, CDFs that allow the nuclear saturation 
characteristics in the isoscalar channel, including the skewness 
coefficient $Q_{\rm sat}$, to be freely adjusted provide a robust 
framework for modeling nuclear and CS matter. Our Bayesian analysis 
further shows that different functional forms in the isovector channel 
yield comparable posterior results for the bulk properties of CSs, 
but lead to notable differences in the density dependence of the 
symmetry energy and, correspondingly, in the particle composition of 
dense matter. This underscores that probing the nuclear symmetry energy 
at high densities requires alternative strategies, such as studying 
CS cooling. The rational-function parametrization of the isovector meson 
coupling introduced here incorporates two independent parameters, 
enabling flexible tuning of nuclear saturation properties, such as the 
curvature coefficient $K_{\rm sym}$. This approach provides a suitable 
framework for future simulations of CSs and heavy-ion collisions aimed 
at probing the effects due to isospin asymmetry.

\section*{Acknowledgments}
G.~W. and J.~L. acknowledge the support of the National Natural Science
Foundation of China under Grant No. 12475150. A.~S. is supported by the 
Deutsche Forschungsgemeinschaft (DFG) under Grant No.~SE~1836/6-1 and by 
the Polish National Science Centre (NCN) under Grant No.~2023/51/B/ST9/02798. 
Y.~W. and Q.~L. are supported by the National Natural Science Foundation 
of China (Grant Nos. 12335008 and 12505143) and the National Key Research and 
Development Program of China (Grant Nos. 2023YFA1606402 and 2022YFE0103400).


%

\end{document}